\pdfoutput=1
\documentclass[aps,prd,reprint,showpacs,showkeys,preprintnumbers,nofootinbib,onecolumn]{revtex4-2}

\usepackage{graphicx,epsfig,color}
\usepackage[english]{babel}
\usepackage{amssymb,amsfonts,amsmath}
\usepackage{verbatim}
\usepackage{tikz}
\usepackage{bbold,bm,bbm}
\usepackage{hyperref}
\usepackage{tensor}

\def\be{\begin{equation}}
\def\ee{\end{equation}}
\def\ba{\begin{eqnarray}}
\def\ea{\end{eqnarray}}
\def\nb{\nonumber}
\def\p{\partial}
\def\te{\tensor}
 
\def\a{\alpha}
\def\b{\beta}

\def\g{\gamma}

\def\d{\delta}

\def\k{\kappa}
\def\l{\lambda}
\def\L{\Lambda}
\def\m{\mu}
\def\n{\nu}

\def\s{\sigma}
\def\S{\Sigma}

\def\mc{\mathcal}

  \newcommand{\cF}{{\cal F}}

\newcommand{\bea}{\begin{eqnarray}} \newcommand{\eea}{\end{eqnarray}}
\newcommand{\beann}{\begin{eqnarray*}}  \newcommand{\eeann}{\end{eqnarray*}}
\newcommand{\bfig}{\begin{figure}} \newcommand{\efig}{\end{figure}}
\newcommand{\bcen}{\begin{center}} \newcommand{\ecen}{\end{center}}
\newcommand{\btab}{\begin{tabular}} \newcommand{\etab}{\end{tabular}}

\newtheorem{Proposition}{Proposition}[section]

\newtheorem{Theorem}{Theorem}[section]
\newtheorem{Lemma}{Lemma}[section]
\newtheorem{Corrolary}{Corrolary}[section]

\newcommand{\bp}{\begin{Proposition}}   \newcommand{\ep}{\end{Proposition}}
\newcommand{\bt}{\begin{Theorem}}   \newcommand{\et}{\end{Theorem}}
\newcommand{\bl}{\begin{Lemma}}     \newcommand{\el}{\end{Lemma}}
\newcommand{\bc}{\begin{Corrolary}} \newcommand{\ec}{\end{Corrolary}}

\def\ep{\epsilon}

\def\d{\delta}

\def\a{\alpha}
\def\b{\beta}
\def\s{\sigma}
\def\m{\mu}
\def\n{\nu}

\begin{document}

\title{Emergent Dipole Gauge Fields and Fractons}

\author{Alessio Caddeo}
\email{caddeoalessio.uo@uniovi.es}
\affiliation{Department of Physics and Instituto de Ciencias y Tecnolog\'{\i}as Espaciales de Asturias (ICTEA) \\ Universidad de Oviedo, c/ Federico Garc\'{\i}a Lorca 18, ES-33007 Oviedo, Spain}

\author{Carlos Hoyos}
\email{hoyoscarlos@uniovi.es}
\affiliation{Department of Physics and Instituto de Ciencias y Tecnolog\'{\i}as Espaciales de Asturias (ICTEA) \\ Universidad de Oviedo, c/ Federico Garc\'{\i}a Lorca 18, ES-33007 Oviedo, Spain}

\author{Daniele Musso}
\email{mussodaniele@uniovi.es}
\affiliation{Department of Physics and Instituto de Ciencias y Tecnolog\'{\i}as Espaciales de Asturias (ICTEA) \\ Universidad de Oviedo, c/ Federico Garc\'{\i}a Lorca 18, ES-33007 Oviedo, Spain}

\begin{abstract}

We present a realization of fracton-elasticity duality purely formulated in terms of ordinary gauge fields, encompassing standard elasticity and incommensurate crystals as those describing twisted bilayer graphene, quasicrystals or more general moir\'e lattices. Our construction comprises a description of all types of two-dimensional defects: disclinations, dislocations, discompressions and point-like defects, and takes into account body forces and impurities. The original form of the duality in terms of tensor gauge fields is recovered after partial gauge fixing. We identify the coupling of each type of defect to the dual gauge fields, and from gauge invariance we derive generalized continuity equations for the defect currents and the expected mobility restrictions of elasticity defects.

\end{abstract}

\maketitle

\section{Introduction}

In $2+1$-dimensional condensed matter systems, superfluids and superconductors admit descriptions in terms of emergent dynamical gauge fields through particle-vortex duality, which maps the current to the field strength of the emergent fields. Superfluid defects, \emph{i.e.} vortices, map to charged particles for the emergent gauge fields \cite{Dasgupta:1981zz,FisherLee}. More recently, it has been shown that a generalization of particle-vortex duality for elasticity and related theories maps the stress tensor to the field strength of tensor gauge fields \cite{Pretko:2017kvd}, based on earlier work by Kleinert \cite{kleinert1982duality,kleinert1983double} (for reviews see \cite{Kleinert,Beekman2017rev}). This has been extended to related systems, including Cosserat elasticity\cite{Gromov:2019waa,Hermele:2020,Hirono:2021lmd}, elasticity with smectic anisotropy \cite{Radzihovsky_Smectic}, supersolids \cite{Pretko:2018,Hirono:2021lmd}, vortex lattices \cite{Nguyen:2020yve}, elasticity of quasicrystals \cite{Surowka:2021ved} and elasticity with underlying moir\'e lattices \cite{Gaa}. In addition, folding and tearing can have an interpretation as being fractonic \cite{Manoj:2020bcz} (see \cite{Grosvenor:2021hkn} for a review).

In the elasticity duals, some defects (disclinations for ordinary elasticity) also map to charged particles, with the additional feature that the dipole moment is conserved too, thus constraining the charges to immobility. Since this is the expected behavior for fracton excitations, the map was dubbed fracton-elasticity duality. A prominent aspect of the duality is that, due to the conserved dipole symmetry, the spatial components of the dual fields form a symmetric rank-two tensor coupled to a corresponding current. Generalizations to higher multipole charges lead to the conservation of higher-rank tensor currents and gauge fields \cite{Pretko:2016kxt,Pretko:2016lgv,Gromov:2018nbv}. Although tensor fields realize the symmetries naturally, there are some standing issues such as the coupling to a curved background geometry (see e.g. \cite{Slagle:2018kqf,Jain:2021ibh,Bidussi:2021nmp}). 

A promising avenue to circumvent the problems of tensor gauge fields consists in implementing multipole symmetries through ordinary vector gauge fields, as in \cite{Pena-Benitez:2021ipo}. In this realization, the generators of multipole transformations and spatial translations are internal symmetries. The origin of these symmetries can be understood from the field theory description of solid materials and fluids \cite{Leutwyler:1996er,Son:2005ak,Dubovsky:2011sj,Nicolis:2013lma,Nicolis:2015sra,Watanabe:2013iia,Moroz:2018noc}. The long-wavelength properties of the material can be captured by an effective action for a set of fields, the material coordinates, describing a map between an element of the material in space and its position in a frame attached to the material. Multipole symmetries are then realized as transformations acting on the target space of material coordinates.

In this paper we propose a formulation of fracton-elasticity duality based on vector gauge fields, where the dipole and the second moment of the charge are conserved. We derive the dual elasticity action, for ordinary elasticity and incommensurate crystals like moir\'e lattices or quasicrystals, and the currents coupled to the dual gauge fields. We show that the (non-)conservation of the currents reproduce the expected physics for elasticity and its defects and re-derive the continuity equation with a tensor current. In addition, we show that the mobility constraints for elastic defects follow from gauge-invariance of the coupling of particles to the dual fields.

\section{Gauging of Dipole Symmetry}

In the effective description of elastic defects the relevant conserved charges are the spatial momenta and the number of each type of defect, namely disclinations, dislocations and point-like defects, vacancies and interstitials. To each of these we associate respectively the generators of spatial translations $P_a$, an Abelian charge $Q_0$, and the first (or dipole) and second spatial moments of the charge $Q_1^a$ and $Q_2$\footnote{Here and in the following, $a,b,\dots$ denote material frame indices, $\mu,\nu,\dots$ spacetime indices and $i,j,\dots$ spatial indices.}.

The symmetry generators obey a non-trivial algebra, with non-vanishing commutators
\begin{equation}\label{PQtQ}
i[P_a,Q_1^b]=\delta_a^b Q_0,\qquad i[P_a,Q_2]=\delta_{ab}Q_1^b\,.
\end{equation}
In principle, the algebra can be extended by adding spatial rotations, with $P_a$ and $Q_1^a$ transforming as vectors.

The symmetry associated to this algebra can be gauged by introducing a connection that contains a gauge field for each of the generators
\begin{equation}\label{con}
    {\cal A}_\mu 
    =
    V_\m ^a P_a
    + a_\mu Q_0  
    + b_{\mu a} Q_1^a 
    + c_{\mu} Q_2\ . 
\end{equation}
The covariant field strength is defined as usual
\begin{equation}\label{curva}
    {\cal F}_{\mu\nu} 
    = 
    \partial_\mu {\cal A_\nu}
    - \partial_\nu {\cal A_\mu}
    + i\left[{\cal A}_\mu,{\cal A}_\nu\right]\ .
\end{equation}
Thus, under an infinitesimal gauge transformation, 
\be
\label{genericgaugetransf}
\L = \kappa^a P_a  + \l_0 Q  +\l_{1\,a} Q_1^a+\l_2 Q_2 \ ,
\ee
the gauge connection  \eqref{con} and its field strength \eqref{curva} transform as expected
\ba
\label{gaugetransA}
\d \mc{A}_\m &=& \p_\m \L + i[\mc{A}_\m, \L] \ , \\
\d \mc{F}_{\m \n} &=& i[\mc{F}_{\m \n}, \L] \ .
\ea
We identify the components of the field strength as follows
\begin{equation}
    {\cal F}_{\mu\nu}
    =
    {\cal V}_{\mu\nu}^a P_a
	+{\cal F}_{\mu\nu}^{(0)} Q_0    
    + {\cal F}_{\mu\nu\,a}^{(1)} Q_1^a
	+{\cal F}_{\mu\nu}^{(2)} Q_2 \ ,
\end{equation}
where
\begin{subequations}
\ba
& \mc{V} _{\m \n}^{a} =& \p_\m V^a _\n - \p_\n V^a _\m \, , \\
& \mc{F} _{\m \n} ^{(0)} =& \p_\m a_\n - \p_\n a_\m +   V_\m ^a  b_{\n a} - V_\n ^a   b_{\m a} \, ,\\
& \mc{F} _{\m \n\, a}^{(1)} =& \p_\m b_{\n a} - \p_\n b_{\m a}+\delta_{ab}\left( V_\m ^b  c_{\n} - V_\n ^b   c_{\m}\right)\,, \\  \label{FQ}
& \mc{F} _{\m \n} ^{(2)} =& \p_\m c_\n - \p_\n c_\m  \, .
\ea
\end{subequations}

Under the gauge transformation \eqref{gaugetransA}, the gauge potentials change according to
\begin{subequations}
\label{transfconnection}
\ba
&\d V_\m ^a =& \p_\m \k^a  \, , \\
&\d a_\m  =& \p_\m \l_0 +  V_\m ^a   \l_{1\,a} - b_{\m a} \k^a \,, \\
&\d b_{\m a} =& \p_\m \l_{1\,a}+\delta_{ab}\left(V_\mu^b \l_2-c_\mu \k^b\right) \, , \\
&\d c_\m  = & \p_\m \l_2\ .
\ea
\end{subequations}
The field strengths $\mc{V} _{\m \n}^{a}$ and $\mc{F} _{\m \n}^{(2)}$ are invariant, but the other components transform non-trivially
\ba
\label{Qcurvaturetransformation}
&\d \mc{F}_{\m \n} ^{(0)} =& \mc{V}_{\m \n} ^{a} \l_{1\,a}   -  \mc{F}_{\m \n\, a} ^{(1)} \kappa^a \,,\\
&\d \mc{F}_{\m \n\,a} ^{(1)} =& \d_{ab} \left(\mc{V}_{\m \n} ^{b} \l_2  -   \mc{F}_{\m \n} ^{(2)} \kappa^b\right)   \, .
\ea

\section{Fracton-Elasticity Duality}

As mentioned before, the effective long-wavelength dynamics of a material is captured by the material coordinates $X^a$. For a crystal solid or similar materials, the material translations shifting $X^a\to X^a+\kappa^a$ and the spatial translations shifting the coordinates $x^i\to x^i+k^i$ are spontaneously broken to a diagonal subgroup by a background value of the fields. The fluctuations around this background constitute the displacement fields $u^a$
\begin{equation}
X^a(t,\bm{x})=\d^a_i x^i+u^a(t,\bm{x})\ . 
\end{equation}
The material coordinates can be coupled to an external gauge field for translations $v_\mu^a$, in such a way that the effective action should be invariant under local material translations. This can be achieved through a covariant derivative, that for the background value of the material coordinates will have the value
\begin{equation}
D_\mu X^a=\partial_\mu X^a-v_\mu^a=\d_\mu^a+\partial_\mu u^a-v_\mu^a\ .
\end{equation}
Therefore, an equivalent formulation of the effective theory corresponds to expanding around a trivial background $X^a=u^a$ with a non-vanishing external field $V_\mu^a=-\delta_\mu^a+v_\mu^a$. We henceforth set other external sources to zero $v_\mu^a=0$, but we emphasize that the effective theory could in principle be maintained invariant under local material translations if a general external field is kept. The background value of $V_\mu^a$ also allows us to trade spatial and material frame indices, which we will do occasionally without further comment.

The low-energy effective action for the displacement fields, expanded to lowest order in derivatives and amplitudes, has the form 
\begin{equation}\label{eq:elasticaction}
\mc{L}=\frac{1}{2} \delta_{ab}\dot{u}^a\dot{u}^b-\frac{1}{2} \te{C}{^i _a ^ j _b} \partial_i u^a\partial_j u^b+f_a u^a\,, 
\end{equation}
where $f_a$ includes an intrinsic part corresponding to elastic point-like defects and an external part corresponding to other non-elastic defects, such as impurities, and to body forces.

We henceforth restrict the analysis to two-dimensional systems. Depending on the characteristics of the tensor of coefficients $\te{C}{^i _a ^ j _b} $, we may be describing different types of systems. If it is symmetric under permutations within each pair of indices $i\leftrightarrow a$, $j\leftrightarrow b$, then the action describes ordinary elasticity with $u^a$ the displacement fields. If instead there is an antisymmetric term $\sim \te{\epsilon}{^i _a}\te{\epsilon}{^j _b}$, then the action describes the dynamics of phasons in incommensurate crystals such as twisted bilayers or quasicrystals \cite{Gaa,Surowka:2021ved}. There could also be other cases where the trace or shear components of $C$ vanish, but we do not consider them here.

In order to implement the duality, one introduces auxiliary fields corresponding to the spatial momentum density and stress tensor $\pi_a$, $\te{T}{ _a ^i}$
\be
\label{dualelasticityaction}
\begin{split}
 \mc{L} =  &
 -\frac{1}{2} \delta^{ab}\, \pi_a  \pi_b 
 +
 \frac{1}{2} \te{(C^{-1})}{_i ^a _ j ^b}\, \te{T}{_a ^i} \te{T}{_b ^j}+f_a u^a\\ 
 &+\pi_a  \p_t u^a
 + \te{T}{_a ^i}  \p_ {i} u^a
 -\frac{\sigma}{2} \te{\epsilon}{_i ^a}\te{\epsilon}{^j _b} \te{T}{_a ^i}\p_j u^b\ .
\end{split} 
 \ee
In the case of phasons $\sigma=0$, while for ordinary elasticity $\sigma=1$. Solving the equations of motion for the auxiliary fields results in the original action \eqref{eq:elasticaction}. On the other hand, the equations of motion for $u^a$ become
\begin{equation}\label{eq:Tcons}
\dot{\pi}_a+\partial_i \left( \te{T}{_a ^i}-\frac{\sigma}{2} \te{\epsilon}{_a ^i}\te{\epsilon}{^b _j} \te{T}{_b ^j}\right)=f_a\,.
\end{equation}
For incommensurate crystals ($\sigma=0$) these equations coincide with the conservation of the momentum currents $\partial_\mu\te{T}{_a ^\mu}$, where $\te{T}{_a ^t}\equiv \pi_a$. For ordinary elasticity, instead, only the symmetric part of the stress tensor $\te{T}{_a ^i}$ enters the conservation equation. This difference will be reflected in the parametrization of the momentum-stress components in terms of the dual gauge fields.

First, we split the fields $u^a = u^{(s)\,a} + \tilde u^a $ in a singular part $u^{(s)\,a}$ containing the defects and a smooth part $\tilde u^a$.

After this splitting, the smooth part is integrated out, leaving an action for the auxiliary fields of the form \eqref{dualelasticityaction} whith $u^a\to u^{(s)\,a}$ plus a constraint implementing the conservation of the energy-momentum tensor. The dual fields are introduced as a parametrization of the energy-momentum tensor that satisfies the constraint \eqref{eq:Tcons} automatically, this is achieved by the ansatz
\ba\label{eq:dualTall}
&\pi_a= &\te{\epsilon}{_a ^b}\left(B_b^{(1)}-\left(1-\frac{\sigma}{2}\right)B_b^{(0)}\right)\,, \label{eq:dualT} \\ 
&\te{T}{_a ^i}= &-\te{\epsilon}{_a ^b}\epsilon^{ij}\left(E_{jb}^{(1)}-E_{jb}^{(0)}\right)\,, \label{eq:dualT2}
\ea
where the dual electric and magnetic fields are defined as follows:
\begin{align}
\label{eE}
& E_{ia}^{(1)} = \mc{F}_{t i\,a} ^{(1)} \ , & E_{ia}^{(0)} &= \partial_t\mc{F}^{(0)} _{ia}-\partial_i \mc{F}^{(0)} _{ta}\ , \\
\label{bB}
& B_a^{(1)} = \frac{1}{2} \epsilon^{ij}\mc{F}_{ij\,a} ^{(1)} \ ,  & B_a^{(0)} &= \epsilon^{ij} \partial_i\mc{F}^{(0)}_{ja} \ .
\end{align}
The conservation equation \eqref{eq:Tcons} with the dual fields results in
\begin{equation}\label{eq:constrc}
E_i^{(2)}\equiv \mc{F}^{(2)}_{t i}=f_i\,. 
\end{equation}
At the level of the effective action this constraint can be implemented through a Lagrange multiplier. 
In the absence of external forces or point-like defects, one can set $c_\mu=0$.  An analysis of the fluctuation spectrum reveals that the dual theory reproduces the dispersion relations of phonons and phasons. Details can be found in the Appendix. For ordinary elasticity one might further force a constraint that removes the antisymmetric part of the the stress tensor $\epsilon^{ij}T_{ij}=0$, this leads to the condition
\begin{equation}\label{eq:fracconst}
0=\te{\epsilon}{^i ^j}\left(E_{ij}^{(1)}-E_{ij}^{(0)}\right)=-\te{\epsilon}{^i^j}\partial_t\left(\partial_i a_j+b_{ij}\right)\ .
\end{equation}
This constraint is satisfied for
\begin{equation}
b_{ij}=b_{(ij)}+\partial_i \lambda_j, \ a_i=\partial_i \alpha-\lambda_i,\ \ a_t=\phi+\partial_t\alpha\ .
\end{equation}
The gauge can be partially fixed by demanding that $b_{ij}$ is symmetric and $a_i=0$, in which case the set of allowed gauge transformations is restricted to $\lambda_i=\partial_i \beta$, $\alpha=\beta$.  One then recovers the tensor fields and gauge transformations of the original formulation of fracton-elasticity duality.

\subsection{Global Dipole Symmetry and Defect Currents}

In the dual formulation there are emergent global symmetries corresponding to the gauge transformations of the charge and the moment fields \eqref{transfconnection} that leave the gauge potentials invariant. They can be identified according to the moment of the charge density that is conserved.
\begin{center}
\begin{tabular}{l|ccc}
 moment & $\l_0$ & $\l_{1\,a}$ & $\l_2$ \\ \hline
 0th & \ const \ & \ 0 \ & \ 0 \ \\
 1st & \ $x^i\l_{1\,i}$ \ & \ const${}_a$ \ & \ 0\ \\
 2nd & \ $\frac{\bm{x}^2}{2}\l_2$ \ & \ $\delta_{ai}x^i\l_2$ \ & \ const \
\end{tabular}
\end{center}
For each of the global symmetries there is a conserved current. We can derive the currents by identifying the coupling of the gauge potentials to the elasticity defects and taking into account the gauge invariance of the emergent fields.

After implementing the duality we are left with a quadratic action for the dual gauge fields plus a coupling to the singular parts of the displacement/phason fields, the second line in \eqref{dualelasticityaction} with $u^a$ replaced by the singular part $u^{(s)\,a}$ and the momentum density and stress tensor replaced by their expressions in \eqref{eq:dualT}, \eqref{eq:dualT2}.

Expanding and integrating by parts, this part of the action can be cast as
\begin{equation}\label{eq:LJ}
\mc{L}_J=-a_\mu J^{(0)\,\mu}-b_{\mu a} J^{(1)\,\mu a} -c_\mu J^{(2)\,\mu}\,.
\end{equation}
These currents will be related to the currents for disclinations $\Theta^\mu=\epsilon^{\mu\nu\lambda}\partial_\nu\partial_\lambda\theta^{(s)}$, where $\theta^{(s)}=\epsilon^{ij}\partial_i u_j^{(s)}/2$, and dislocations $D^{\mu a}=\epsilon^{\mu\nu\lambda}\partial_\nu\partial_\lambda u^{(s)\,a}$. For linear elasticity they combine in the defect current \cite{Kleinert} $j^{\mu a}=\epsilon^{ab}\epsilon^{\mu\nu\lambda}\partial_\nu u_{\lambda b}^{(s)}$, where $u_{ta}^{(s)}=\partial_t u_a^{(s)}/2$ and $u^{(s)}_{ij}=(\partial_i u^{(s)}_j+\partial_j u^{(s)}_i)/2$.
Even in the absence of boundaries, the action \eqref{eq:LJ} is gauge-invariant only under certain conditions that elasticity defects must satisfy, otherwise one cannot discard the total derivative terms that appear upon integration by parts. Following \cite{Kleinert}, the singular displacement is a multi-valued field that is analog to a singular diffeomorphism producing a Riemann curvature
\begin{equation}
R_{\mu\nu,\alpha\beta}=(\partial_\mu \partial_\nu-\partial_\nu \partial_\mu)\partial_\alpha u^{(s)}_\beta.
\end{equation}
Where $u^{(s)}_t=0$. For elasticity the condition $R_{\mu\nu,(ij)}=0$ is satisfied, implying that  $u_{ij}^{(s)}$ single-valued and the same applies to $\partial_\mu\theta^{(s)}$. For $\sigma=1$ gauge invariance of \eqref{eq:LJ} needs in addition that $R_{\mu\nu,ti}=0$, such that $\partial_t u_i^{(s)}$ is single-valued. Geometrically this would mean that vectors only change by spatial rotations under parallel transport. For $\sigma=0$ there is one more condition that is needed $R_{ti,kl}=0$, such that parallel transport along the time direction is trivial. The last condition forces the spatial $\Theta^i=0$ to vanish. It might be possible to relax these conditions if the total derivative terms are considered, but we will not pursue this here. 

The currents coupling directly to the individual gauge fields are 
\begin{subequations}\label{eq:defectcurrents}
\ba
&J^{(2)\,t}  = \partial_i u^{(s)\,i}\,,\ &J^{(2)\,i} = -\partial_t u^{(s)\,i}\,,  \\
&\!\!\!\!\!\! J^{(1)\,ta} =  0\,, \ \ \ \  &J^{(1)\,ia} =j^{ai}+\frac{1}{2}\epsilon^{ij}\p_j\left(\epsilon^{ab}\p_t u_b^{(s)}\right)+\frac{1-\sigma}{2}\epsilon^{ia}D^b_{\ b}\,,\\
&J^{(0)\,t} = -\partial_a j^{t a}=-\Theta^t-\partial_a D^{t a}\,, \ \ \  &J^{(0)\, i} = -\partial_a j^{ia}+\frac{1-\sigma}{2}\epsilon^{ij}\p_j D^a_{\ a}\,.
\ea
\end{subequations}

From the gauge transformations \eqref{transfconnection} we obtain the continuity equations
\begin{subequations}\label{eq:ward}
\ba
\partial_\mu J^{(0)\,\mu} &=& 0\,,\label{eq:ward1}\\ 
\partial_\mu J^{(1)\,\mu a} &=& -J^{(0)\,a}\,,\label{eq:ward2}\\
\partial_\mu J^{(2)\,\mu} &= & -\delta_{ia}J^{(1)\,i a}\,.
\ea
\end{subequations}
These reproduce the expected relations between currents for disclinations, dislocations and point-like defects respectively. It can be checked explicitly that \eqref{eq:defectcurrents} satisfy the equations above. The conserved currents associated to each of the global symmetries are
\begin{subequations}\label{eq:globalcurrents}
\ba
\mc{J}_0^\mu &=&  J^{(0)\,\mu}\,, \\ 
\mc{J}_1^{\,\mu a} &=& J^{(1)\,\mu a} +x^a J^{(0)\,\mu}\,,\\
\mc{J}_2^{\mu} &= & J^{(2)\mu} +\delta_{ia}x^i J^{(1)\,\mu a}+\frac{\bm{x}^2}{2} J^{(0)\,\mu}\,.
\ea
\end{subequations}

The defect currents admit improvement terms that are consistent with \eqref{eq:ward}. The improved currents have the general form
\begin{subequations}\label{eq:improvements}
\ba
\widetilde{J}^{(0)\,\mu} &=&  J^{(0)\,\mu}+\partial_\alpha \Psi_1^{\alpha\mu}\,, \\ 
\widetilde{J}^{(1)\,\mu a} &=& J^{(1)\,\mu a} -\Psi_1^{\mu\,a}+\partial_\alpha \Psi_2^{\alpha\mu a} \,,\\
\widetilde{J}^{(2)\,\mu} &= & J^{(2)\mu} -\Psi_2^{\mu\nu a}\delta_{\nu a}\,.
\ea
\end{subequations}
Where $\Psi_1^{\mu\nu}=-\Psi_1^{\nu\mu}$ and $\Psi_2^{\mu\nu a}=-\Psi_2^{\nu\mu a}$.
It is possible to choose the improvement terms in such a way that $\widetilde{J}^{(2)\,\mu}=0$, $\widetilde{J}^{(1)\,ta}=0$, and $\widetilde{J}^{(1)\,ia}$ is symmetric and traceless. With the improved currents, the full set of continuity equations can be reduced to the usual continuity equation for a tensor current with the tracelessness condition
\begin{equation}\label{eq:constens}
\partial_t \widetilde{J}^{(0)\,t}-\partial_i \partial_j \widetilde{J}^{(1)\,ij}=0,\qquad \widetilde{J}^{(1)\,i}_{\phantom{(1)}\,\ i}=0\,.
\end{equation}
On the other hand, the improvements with $\Psi_1^{a t}=-\Psi_1^{t a}=D^{t a}$ make the defect densities $\widetilde{J}^{(0)\,t}$ and $\widetilde{J}^{(1)\,t}$ coincide with the usual definitions of disclination and dislocation densities.

\section{Fractons from Dipole Gauge Invariance}

According to standard arguments of dipole conservation, a particle charged under the charge $Q$ should be immobile. This can be understood from the gauged dipole symmetry. A charged particle of charge $q$ following a trajectory $x^\mu(s)$ through spacetime couples to the gauge field component $a_\m$ through a term proportional to
\be
S_q=q\int_\g a \equiv q\int_\g d s \, \dot{x} ^\m (s)\,  a_\m \ ,
\ee
where $\g$ is the particle worldline. This is invariant under the monopole gauge transformation in \eqref{transfconnection}, but
in general it is not invariant under dipole transformations. In fact, the dipole variation is
\be
\d S_q=q\d \int_\g a = -q\int _\g d s \,  \dot{x} ^a (s) \, \l_a\left[x(s)\right]  \ .
\ee
For a constant $\l_a$ the variation is proportional to the spatial vector connecting the two endpoints of the trajectory, \emph{i.e.} the change in dipole moment generated by the motion of the particle. The variation vanishes for all gauge transformations only if $\g$ is extended in the time direction. This is precisely the trajectory of a particle that stays at a fixed point in space, so dipole gauge invariance implies that charges are immobile fractons.

The conservation of the second moment of the charge imposes analogous constraints on the mobility of a dipole. In this case the motion is restricted to the direction transverse to the dipole moment, that for a dislocation defect in elasticity coincides with the direction of the Burgers vector. A particle with dipole charge $d^a$ couples to the dual gauge field as follows
\begin{equation}\label{eq:dipoleact}
S_{\bm{d}}=d^a\int_\g  A^{(1)}_a=d^a\int_\g ds \, \dot{x}^\m(s) \,  A^{(1)}_{\mu a}\ ,
\end{equation}
where the potential corresponds to the combination of electric fields in the stress tensor \eqref{eq:dualT2} and \eqref{eE}
\begin{equation}
A^{(1)}_{\mu a}=b_{\mu a}-\delta_a^\nu\mc{F}^{(0)}_{\mu\nu}\ .
\end{equation}
With this choice, the action for a static particle becomes
\begin{equation}
S_{\bm{d}}=\int dt \, d^i\partial_i a_t\ .
\end{equation}
This is consistent with the formulas we derived in \eqref{eq:defectcurrents} and \eqref{eq:ward}, the density $J^{(1)\, t a}$ does not receive any contribution, yet the particle is a point-like dipole source for the zeroth moment gauge field.

The action of the dipole particle is invariant under the zeroth and first moment gauge transformations in \eqref{transfconnection}, but not under general second moment transformations
\begin{equation}
\d S_{\bm{d}}=d^a\d \int_\g  A_a ^{(1)}=-d^a\int_\g ds \, \dot{x}^\m(s) \,  \lambda_2[x(s)] \delta_{\mu a}\,.
\end{equation}
The variation vanishes for an arbitrary gauge transformation only if it satisfies the constraint $\bm{d}\cdot \bm{\dot{x}}=0$, which precisely corresponds to the condition of motion in the direction transverse to the dipole moment. 

Consider now particles carrying second moment charge, with a coupling to the dual gauge fields
\begin{equation}
S_{q_2}=q_2\int_\g  A^{(2)} =q_2\int ds\, \dot{x}^\mu(s) A^{(2)}_\mu.
\end{equation}
In principle $A^{(2)}_\mu=c_\mu$ is a possible gauge-invariant coupling, but it produces a contribution to the second moment current that is conserved, so it would correspond to a non-elastic point-like defect. Instead, the form of the ``potential''  should be
\begin{equation}
A^{(2)}_\mu=\frac{1}{2}\delta^{\nu a}\left(\mc{F}^{(1)}_{\mu\nu a}-\partial_\mu \mc{F}^{(0)}_{\nu a}+\partial_\nu \mc{F}^{(0)}_{\mu a}\right).
\end{equation}
Since it only involves the field strenghts, it is gauge-invariant by construction. Evaluating the action for a static particle shows explicitly that the particle acts as a point-like source for the second moment of the charge
\begin{equation}
S_{q_2}=q_2\int dt\, \left( c_t+\frac{1}{2} \bm{\partial}^2 a_t\right).
\end{equation}

In addition to these defects, in an incommensurate crystal there are defects corresponding to the endpoints of strings of dislocations with Burgers vector pointing along the string, identified in \cite{Gaa} as {\em discompressions}. Denoting by $X^\mu(S)$ the embedding functions of the string and $S^\alpha$, $\alpha=0,1$ the worldsheet coordinates, the coupling we propose is
\be\label{eq:actdiscomp}
S_{\bm{b}} = b\int_\S  \cF^{(0)} \equiv b\int _\S d^2 S\, \epsilon^{\a \b} \p_\a X^\m \p_\b X^\n\delta_{\n a}\epsilon^{ab} A_{\m b} ^{(1)} \ ,
\ee
where $\S$ is the worldsheet and $b$ is the modulus of the Burgers vector $b^i=b\hat{b}^i$. For a static string extended along the Burgers vector $X^0=S^0\equiv t$, $X^i=\hat{b}^i S^1\equiv \hat{b}^i s$, the action for a discompression located at the origin of space is
\be
S_{\bm{b}} =  \int dt \int_{-\infty}^0 ds\, \epsilon_i^{\ j} b^i \partial_j a_t \,.
\ee
This shows that indeed there is a density of dipole moment $d^i=\epsilon_i^{\ j} b^j$ pointing in the direction transverse to the string. Under a dipole gauge transformation
\be
\delta S_{\bm{b}}=b\int d^2 S \epsilon^{\a \b} \p_\a X^\m \p_\b X^\n\delta_{\n a}\epsilon^{ab} \partial_\m \lambda_b[X(S)].
\ee
Gauge invariance requires that $\partial_\tau X^i=0$ at the endpoint. Therefore, a discompression is immobile. Furthermore, under a second moment gauge transformation the action changes as
\be
\delta S_{\bm{b}}=b\int d^2 S \epsilon_{ij}\epsilon^{\a \b} \partial_\a X^i \partial_\b X^j \lambda_2[X(S)].
\ee
Now gauge invariance imposes the constraint $\epsilon_{ij}\partial_\tau X^i \partial_\sigma X^j=0$. This condition fixes the transverse velocity of a string element to vanish.

\section{Discussion}

The realization through the dipole gauge symmetry of fracton-elasticity duality unveils a more general set of continuity equations \eqref{eq:ward}, with an independent current associated to each type of defect. The continuity equation with a tensor current \eqref{eq:constens} that captures charge and dipole conservation can be derived by combining the more general equations. The defect currents are not all conserved, nonetheless there are conserved currents associated to the global charge, dipole and second moment symmetries \eqref{eq:globalcurrents}. The global currents contain orbital and intrinsic contributions, in such a way that charges can move by the emission/absorption of dipoles, and dipoles can move along the dipole direction by the emission/absorption of second moment charges. Furthermore, the coupling of particles to the dipole gauge fields confirms the mobility restriction of isolated defects. Thus all the expected mobility features of elasticity defects are derived naturally from dipole gauge symmetry. In principle it should be possible to implement a similar realization of fracton-elasticity duality with emerging dipole gauge symmetry to other theories related with more degrees of freedom, such as Cosserat elasticity, supersolids, \emph{etc}.

For an incommensurate crystal it is possible to change the relative coefficient between the zeroth moment and the first moment fields in both \eqref{eq:dualTall} simultaneously. In particular, for a vanishing coefficient the expressions coincide with those in \cite{Gaa}, with a suitable identification of the gauge potentials. This does not alter the continuity equations, nor the values of the physical momentum and stress, but it modifies the expressions for the defects currents, and the coupling of defects to the gauge fields.

Our derivation of the general continuity equations may have consequences for fracton hydrodynamics, originally formulated in \cite{Gromov:2020yoc} and extended to momentum-preserving systems in \cite{Grosvenor:2021rrt,Osborne:2021mej} and to magnetohydrodynamics in \cite{Qi:2022seq}. Note that by using the improvement terms \eqref{eq:improvements}, one recovers indeed the continuity equation of tensor currents \eqref{eq:constens} that is the starting point of fractonic hydrodynamics. In this respect, the effective hydrodynamic description is unchanged, nevertheless one may consider extensions analogous to the inclusion of a spin current in relativistic hydrodynamics \cite{Hattori:2019lfp,Li:2020eon,Gallegos:2021bzp,Hongo:2021ona,Gallegos:2022jow}.

Another interesting direction would be to explore further the dynamics of individual defects from the point of view of particle or spatially extended defects, in view of the restrictions imposed by dipole gauge invariance and possible higher multipole symmetries. There could also be interesting connections with non-relativistic quantum gravity, as the coupling of the discompression to the first moment gauge field we propose \eqref{eq:actdiscomp} is similar to the coupling of a non-relativistic string action to the $B$-field as implemented in \cite{Harmark:2019upf,Bergshoeff:2019pij,Bidussi:2021ujm}.

\begin{acknowledgments}
{\em  Acknowledgments}$\;$ We thank Francisco Pe\~{n}a-Benitez and Piotr Surowka for useful comments. We also thank Jay Armas and Akash Jain for illuminating discussions. The work of A.C.~is supported by the ``Fondazione Angelo Della Riccia". A.C., C.H. and D.M. are partially supported by the AEI through the Spanish grant PGC2018-096894-B-100 and by FICYT through the Asturian grant SV-PA-21-AYUD/2021/52177.

\end{acknowledgments}

\appendix

\section{Elastic modes}

In this section we write the equations of motion of the dual gauge fields and compare with the modes obtained from the elastic theory in the simple case of a triangular lattice or isotropic system, where the tensor of the elastic coefficients admits the decomposition
\be
\label{triage}
\te{C}{^i _a ^ j _b}
= 
G \te{(P_2)}{^i _a ^ j _b}
+H \te{(P_1)}{^i _a ^ j _b}
+K \te{(P_0)}{^i _a ^ j _b}\ ,
\ee
having introduced the projectors
\begin{align}
    \te{(P_0)}{^i _a ^ j _b}
    & \equiv \frac{1}{2} \delta^{i} _a \delta_{b} ^j\ , \qquad
    \te{(P_1)}{^i _a ^ j _b}
     \equiv \frac{1}{2}(\delta^{ij}\delta_{ab} -\delta^i_b\delta^j_a)\ , \qquad
    \te{(P_2)}{^i _a ^ j _b}
     \equiv \frac{1}{2} (\delta^{ij}\delta_{ab} +\delta^i_b\delta^j_a -\delta^i_a \delta^j_b)\ .
\end{align}
The coefficients are the shear modulus $G$, bulk modulus $K$ and rotational stiffness $H$, with the last vanishing for an ordinary crystal but being non-zero for the incommensurate crystal. The inverse of the elasticity coefficients tensor is
\be
\label{intriage}
\te{(C^{-1})}{^i _a ^ j _b}
= 
\frac{1}{G} \te{(P_2)}{^i _a ^ j _b}
+\frac{1}{H} \te{(P_1)}{^i _a ^ j _b}
+\frac{1}{K} \te{(P_0)}{^i _a ^ j _b}\ ,
\ee
With these definitions of the elastic coefficients, the dispersion relations $\omega(\bm{k})$ of the phonons/phasons are
\be\label{eq:dispersion}
\omega^2=\frac{G+H(\sigma-1)^2}{2}\bm{k}^2,\qquad \omega^2=\frac{G+K}{2}\bm{k}^2,\ 
\ee
for transverse and longitudinal modes, respectively. Note that \eqref{eq:dispersion} accounts for both the $\sigma = 0$ and $\sigma = 1$ cases.

\subsection{Equations of motion}

After introducing the dual fields \eqref{eq:dualTall} in the action \eqref{dualelasticityaction}, we obtain the equations of motion for the dual fields by taking variations with respect to the gauge potentials. For convenience we define
\be
\te{(\widetilde{C}^{-1})}{^{iajb} } = \te{(C^{-1})}{_k ^c _l ^d } \te{\epsilon}{_c ^a }  \te{\epsilon}{_d ^b } \te{\epsilon}{^{i k }} \te{\epsilon}{^{j l}} \ .
\ee
Note that in the isotropic case $\te{(\widetilde{C}^{-1})}{^{iajb} }=\te{(C^{-1})}{^{iajb} }$.
The equations of motion for $a_\mu$ are
\ba
\te{(\widetilde{C}^{-1})}{^{iajb} } \p_i \p_a \left( E^{(1)}_{jb} - E^{(0)}_{jb} \right) &=& - J^{(0)\, t} \ , \\
\left(1 - \frac{\s}{2}  \right) \epsilon^{ij}  \p_j \p_a \left[B^{(1)}_{a} - \left(1- \frac{\s}{2} \right) B^{(0)}_{a} \right] + 
\te{(\widetilde{C}^{-1})}{^{iajb} } \p_t \p_a \left( E^{(1)}_{jb} - E^{(0)}_{jb} \right) &=&  J^{(0)\, i} \ .
\ea
The equation of motion for $b_{t a}$ vanishes (since the current density is zero), while those for $b_{ia}$ read
\ba
 \left[  \frac{\s}{2} \d^{ab} \epsilon^{i j} - \left( 1- \frac{\s}{2} \right) \d^{i b} \epsilon^{ j a} \right] \p_j \left[B^{(1)}_{b} - \left(1- \frac{\s}{2} \right) B^{(0)}_{b} \right] + \nb \\
+ \te{(\widetilde{C}^{-1})}{^{a i j b} } \p_t \left( E^{(1)}_{j b} - E^{(0)}_{j b} \right)  = -  J^{(1)\, i a} \ .
\ea
The equations of motion for $c_\mu$ are
\ba
\d_{ia}  \te{(\widetilde{C}^{-1})}{^{iajb} }  \left( E^{(1)}_{jb} - E^{(0)}_{jb} \right) - \p_ i \phi^ i &=&  J^{(2)\, t} \ , \\
\epsilon^{ia} \left[B^{(1)}_{a} - \left(1- \frac{\s}{2} \right) B^{(0)}_{a} \right] - \p_t \phi ^i &=& - J^{(2)\, i} \ ,
\ea
where $\phi^i$ are the Lagrange multipliers for the constraint \eqref{eq:constrc}, in fact their equations of motion are
\be
E_i^{(2)} = f_i\ .
\ee
One can check that the Ward identities \eqref{eq:ward} are consistent with these set of equations. For vanishing forces and defect currents, the non-trivial solutions to the equations of motion have the expected dispersion relations \eqref{eq:dispersion} and give raise to the same values of momentum and stress as the original elasticity theory.

\subsection{Fields for the mode solutions}

We refer to the elastic dispersion relations \eqref{eq:dispersion} as transverse and longitudinal mode, respectively. 
The field and Lagrange multiplier solution corresponding to the transverse mode is
\begin{align}
    E^{(1)}_{ia} &= 
    G k_i \tilde k_a 
    +\frac{1}{2} \epsilon_{ia}
    \left[G+H(\sigma-1)\right]\bm{k}^2\ ,\qquad
    B^{(1)}_a = \frac{1}{2} \left[ 
    G + H(\sigma-1)\right]
    \frac{k_a \bm{k}^2}{\omega}\ ,\\
    E^{(0)}_{ia} &= \epsilon_{ia} H(\sigma-1) \bm{k}^2\ , \qquad 
    B^{(0)}_a =  H(\sigma-1)\frac{k_a \bm{k}^2}{\omega}\ , \qquad
    \phi_i = i \tilde k_i\ .
\end{align}
where $\tilde k_i \equiv \epsilon_{ij} k_j$.
The field and Lagrange multiplier solution corresponding to the longitudinal mode is
\begin{align}
    E^{(1)}_{ia} &= 
    G k_i k_a 
    -\frac{1}{2}\delta_{ia} \left(G+K\right) \bm{k}^2 \ ,\qquad 
    B^{(1)}_a = -\omega \tilde k_a\ ,\\
    E^{(0)}_{ia} &= 0\ ,\qquad
    B^{(0)}_a = 0\ ,\qquad
    \phi_i = i k_i\ .
\end{align}

\section{Solutions in the presence of defects}

The solution for the magnetic field and electric field for $\phi^i=0$ is
\begin{equation}
    \begin{split}
        B_a^{(1)}-\left(1-\frac{\sigma}{2}\right)B_a^{(0)}= & -\epsilon_{ab}\partial_t u^{(s)\, b},\\
        E_{ia}^{(1)}-E_{ia}^{(0)}=& \epsilon_{ik}\epsilon_{ac} \te{C}{^k^c ^j _b}\partial_j u^{(s)\,b}.
    \end{split}
\end{equation}
In the tensor of elasticity coefficients we have to do the replacement $H\to (1-\sigma)H$ (which does not modify the coefficients for $\sigma=0,1$).
With these solutions, the momentum and stress are
\begin{equation}
    \begin{split}
        \pi_a= & \partial_t u^{(s)\, a},\\
        \te{T}{_a ^i}=&- \te{C}{^i ^a ^j _b}\partial_j u^{(s)\,b}.
    \end{split}
\end{equation}
Combining $B_{ia}=b_{ia}+\partial_i a_a$, a solution is
\begin{equation}
    \begin{split}
        c_t=&\frac{K}{2}\partial_l u^{(s)\,l},\ \ a_t=0, \\
        c_i=&\partial_t u_i^{(s)}+\left(1-\frac{\sigma}{2}\right) \partial_l B_{li}+\frac{\sigma}{2}\partial_l B_{il},\\
        \partial_t B_{ia}=& -\frac{H(1-\sigma)}{2}\epsilon_{ia}\epsilon^j_{\ b}\partial_j u^{(s)\,b}-\frac{G}{2}u_{ia}^{(s)},\\
        u_{ia}^{(s)}=& \partial_i u_a^{(s)}+\partial_a u_i^{(s)}-\delta_{ia}\partial_l u^{(s)\,l}.
    \end{split}
\end{equation}

\bibliographystyle{apsrev4-2}
\bibliography{refs}

\end{document}